\documentclass[prl,amsmath,amssymb,aps,showpacs,10pt,superscriptaddress,twocolumn]{revtex4-1}
\usepackage[utf8]{inputenc}
\usepackage{amsmath}
\usepackage{latexsym}
\usepackage{amsfonts}
\usepackage{epsfig}
\usepackage{psfrag}
\usepackage{graphicx}
\usepackage{hyperref}
\usepackage{fixltx2e}
\usepackage{amssymb,amsmath}
\usepackage{color}
\usepackage{soul,xcolor}
\usepackage[normalem]{ulem}

\newcommand{\bea}{\begin{eqnarray}}
\newcommand{\ea}{\end{eqnarray}}
\newcommand{\eea}{\end{eqnarray}}

\newcommand{\sumint}[1]
{\begin{array}{c} \\
{{\textstyle\sum}\hspace{-1.1em}{\displaystyle\int}}\\
{\scriptstyle{#1}}
\end{array}}

\usepackage[percent]{overpic}

\begin{document}

\preprint{AIP/123-QED}

\title{Mechanisms for carbon adsorption on Au(110)-(2$\times$1): A work function analysis}

\author{H.Z. Jooya}
\affiliation{%
ITAMP, Harvard-Smithsonian Center for Astrophysics, Cambridge, Massachusetts 02138, USA
}%

\author{K.S. McKay}
 \affiliation{%
NIST, 325 Broadway, Boulder, Colorado 80305, USA
}%

\author{E. Kim}
\affiliation{%
Department of Physics and Astronomy, University of Nevada, Las Vegas, Nevada 89154-4002, USA
}%

\author{P.F. Weck}
\affiliation{%
Sandia National Laboratories, P.O. Box 5800, Albuquerque, New Mexico 87185-0779, USA
}%

\author{D.P. Pappas}
\affiliation{%
NIST, 325 Broadway, Boulder, Colorado 80305, USA
}%

\author{D.A. Hite}
\affiliation{%
NIST, 325 Broadway, Boulder, Colorado 80305, USA
}%

\author{H.R. Sadeghpour}
\affiliation{%
ITAMP, Harvard-Smithsonian Center for Astrophysics, Cambridge, Massachusetts 02138, USA
}%

\date{\today}

\begin{abstract}
The  variation of the work function upon carbon adsorption on the reconstructed Au(110) surface is measured experimentally and compared to density functional calculations. The adsorption dynamics is simulated with \textit{ab-initio} molecular dynamics techniques. The contribution of various energetically available adsorption sites on the deposition process is analyzed, and the work function behavior with carbon coverage is explained by the resultant electron charge density distributions.
%
\end{abstract}

\maketitle
\section{Introduction}\label{sec:level1}
Electric field noise from metal surfaces has been a barrier to precision measurements in various miniaturized devices: from the measurement of the Casimir force between near-contact surfaces \cite{Xu2018}, to proof-masses near electrode surfaces in space-based gravitational-wave detectors \cite{Pollack2008};  from spin decoherence for near-surface nitrogen-vacancy (NV) centers in diamond \cite{Kim2015}, to nano-cantilevers probing of dispersion forces \cite{Stipe2001}; and trapped-ion systems \cite{Wineland1998}. 

In ion traps, the electric field noise emanating from the trap electrodes can significantly heat the ions' motion \cite{Brownnutt2015,Kim2017}. This remains a major obstacle to the realization of ion-trap based scalable quantum computing architectures. The surface origin of this noise was supported experimentally upon {\it in situ} treatment of the trap electrodes by ion bombardment, resulting in a reduction in motional heating by more than two-orders of magnitude \cite{Hite2012,Daniilidis2014}, thus validating a microscopic theory of fluctuating surface dipoles as a source of the anomalous field noise \cite{Safavi2011}. Recently, \textit{in-situ} cleaning by ion bombardment has also been employed in a Casimir experiment, reducing detrimental residual potentials by an order of magnitude \cite{Xu2018}. Carbon-bearing adsorbates are suspected to be the dominant contaminants on trap-electrode surfaces \cite{Hite2013,Daniilidis2014}.
When the surface dipoles are mobile and whose magnitude change with motion on the electrode surface, it was shown that the frequency dependence of the noise power spectrum was $\omega^{-3/2}$ and the distance dependence of the spectrum was $d^{-4}$ \cite{Brownnutt2015,Kim2017}, as
\begin{equation}\label{eqn_1}
	S_{E,\perp}\approx\frac{\Delta\mu^2\overline{\sigma}\sqrt{D}}{\sqrt{2}\epsilon_0^2d^4\omega^{3/2}R_P}
\end{equation}
where $\Delta\mu$ is the fluctuation in the induced dipole moment, $R_P$ is the dipole patch radius, $\overline{\sigma}$ is the stationary surface density of adatoms, $D$ is the temperature-dependent diffusion constant, $d$ is the distance of the ion above the surface of the trap electrodes, $\omega$ is the ion motional frequency, and $\epsilon_0$ is the electric permittivity of free space.

Aside from the frequency and distance dependence of the noise power spectrum, it is possible to measure $\Delta\mu$ in surface spectroscopy. In the experiment, $\Delta\mu$ is obtained by measuring the variation of the surface work function caused by adsorption of adatoms, as 
\cite{Jackson1999}
\begin{equation}\label{eqn_2}
\Delta W=\frac{e\Delta \mu}{\epsilon_0 A}
\end{equation}
where $e$ is the electron charge, and $A$ is the surface area taken up by one adatom. The work function, $W$, is the minimum energy needed to remove an electron from the bulk to a point outside of the material \cite{Singh-Miller2009}.

In this work, we extend our earlier density-functional theory (DFT) simulations \cite{Kim2017} to investigate the dependence of the surface noise on the carbon adatom coverage on the gold surface. The work function variation with carbon coverage is calculated and compared with measurement. We study the surface dynamics of carbon adsorbates with {\it ab initio} molecular dynamics \cite{QE2009}; and the contribution of various energetically available adsorption sites on the observed work function is determined.
\section{\label{sec:level1} Methodology}
\subsection{Experimental setup}
The experiments were carried out in an ultra-high-vacuum chamber equipped with instruments for Kelvin probe force microscopy (KPFM), scanning tunneling microscopy (STM), x-ray photoelectron spectroscopy (XPS), and high-resolution low-energy electron diffraction (LEED).  The substrate onto which carbon was deposited was a gold single crystal, cut and polished to within 0.1$^{\circ}$ of the (110) plane.  When cleaned with ion bombardment and annealed to approximately 700 K, the surface displays the usual ($2\times 1$) surface reconstruction with terrace widths on the order of 100 nm with single atomic steps observed in STM, in agreement with sharp ($2\times 1$) LEED spots.  The clean, unannealed Au(110) surface also displays  the ($2\times 1$) reconstructed surface as confirmed by LEED and STM, however terrace widths are  only on the order of 10 nm resulting in broad low-intensity LEED spots.
Each coverage of C/Au(110) in this study was obtained by depositing carbon from a water-cooled carbon sublimation source  onto the clean unannealed Au(110) surface. The deposition rate was around 1 monolayer (ML) per hour. Because carbon sublimates at temperatures near 2500 K, the temperature of the Au(110) reached approximately 500 K during the depositions, which noticeably smoothened the surface topography.

The degree of carbon coverage for each deposition was measured by XPS over an area with an approximate diameter of 200 $\mu$m. The coverage was estimated using a simple model that accounts for the ratio of normalized peak areas for the C 1s and Au 4f$_{7/2}$ core levels and the inelastic mean free path for electrons with those kinetic energies (h$\nu$=1253.6 eV). The relative sensitivity factor between the intensities of core-level photoemission from C and Au was measured for our system using the clean Au(110) and a freshly cleaved sample of highly oriented pyrolytic graphite (HOPG). The definition of 1 ML in this work is equivalent to 2 carbon atoms per 1 Au(110) – ($2\times 1$) unit cell, giving a surface adatom density of 1.7 $g. \mu m^{-2}$. The XPS measurements of the degree of coverage were confirmed by making use of a quartz crystal thickness monitor to measure the surface adatom density of one monolayer to be 1.9(0.6) $g. \mu m^{-2}$. 

Work functions for the various coverages of C/Au(110) up to 3ML were measured with KPFM operated in a frequency modulated mode.  A Cr/Pt-coated Si cantilever with a nominal resonance frequency of 314 kHz was used to measure the contact potential difference (CPD) between the tip and the sample.  In this work, the CPD is defined to be the peak of a Gaussian histogram of a KPFM image with an area of $300 nm \times 300 nm$.  The Gaussian distributions have typical widths on the order of 30 mV FWHM.  To obtain the work function of the C/Au(110) system, CPDs of an HOPG reference sample were measured after each sample coverage to account for any tip changes, and thereby determine the work function of the tip.  The work function of the HOPG reference sample is taken to be 4.6 eV \cite{Takahashi1985}.  For consistency, the work function of the clean unannealed Au(110) was determined to be 5.39 (0.06) eV, as compared to a value taken from the literature of 5.37 eV \cite{Holzl1979}.
\subsection{Computational methods}
First-principles calculations within density-functional theory were carried out using the Plane-Wave Self-Consistent Field (PWSCF) option in the QUANTUM-ESPRESSO (QE)  distribution \cite{QE2009}. The Local-density approximations (LDA) exchange-correlation functional with Perdew and Zunger expression was employed. For gold, the ultrasoft pseudopotential (USPP) was generated with the Rappe Rabe Kaxiras Joannopoulos (rrkjus) scheme with 11 valence electrons in a $5d^{10} 6s$ configuration. For carbon atoms, the $2s^2$ $2p^2$  electrons were treated explicitly as valence electrons in the Kohn-Sham (KS) equations, and the remaining cores were represented by USPP pseudopotentials. A 450 Ry ( 1 Ry=13.605698 eV) kinetic-energy cutoff for the charge density was applied. 
All structures were optimized with periodic boundary conditions applied using the conjugate gradient method, accelerated using the Marzari-Vanderbilt smearing \cite{Marzari1999} with a width of 0.01 Ry. Structural optimizations and property calculations were carried out using the Monkhorst-Pack special k-point scheme \cite{Monkhorst1976} with $6\times 6\times 1$ meshes for integrations in the Brillouin zone (BZ) of the slab systems. 

A ($2\times 2$)-periodic supercell slab was constructed by cleaving relaxed bulk Au with lattice constant 4.14 \text{\AA} , i.e., in close agreement with the experimental value of 4.0780 \text{\AA} at 25$^{\circ}$ \cite{Dutta1963}. The slab model consisted of a six-layer thick Au(110) with the reconstructed ($2\times 1$) superstructure. The ($2\times 1$) reconstruction on Au(110) is called the “missing-row” structure because every second row of the (110) surface chains is missing, as observed in STM (see Fig.~\ref{fig:STM}(a)). 
The top four layers, on the side of the slab used to model atom adsorption, were allowed to relax, while the bottom two layers were kept fixed to mimic the bulk structure. Although a large vacuum region (15\text{\AA}) was used between periodic slabs, the creation of dipoles upon adsorption of atoms on only one side of the slab can lead to spurious interactions between the dipoles of successive slabs. In order to circumvent this problem, a dipole correction was applied by means of a dipole layer placed in the vacuum region following the method outlined by Bengtsson \cite{Bengtsson1999}. As demonstrated in our previous work \cite{Safavi2013,Kim2017}, the introduction of this artificial dipole layer in the vacuum region does not modify the local potential near the surface where adsorption occurs. 

The adsorption energy of carbon on Au(110)-($2\times 1$) surface is measured as $E_{ads}=E_{surf+C}-E_{surf}-E_{C}$, where $E_{surf+C}$, $E_{surf}$, and $E_C$ are the total energies of the surface with a carbon adatom, of the bare surface. and of an isolated C atom, respectively.
Constant-temperature density-functional molecular dynamics simulations (MD) are also carried out using the MD module of the QE package to study the adsorption dynamics of carbon adatoms on the surface. The simulations are done at 300 K to mimic the experimental condition. Fixed surface approximation are found to be an appropriate approach for our qualitative analysis purpose.
\section{Results and Discussion}
It is known that the work function depends strongly on surface orientation, adsorption sites and coverage \cite{Fall1998,Landmann2009,Rhead1988}. The study of work function changes due to the deposition of different species on a solid surface is one of the most promising methods to understand the electronic structure of the surface. 

\begin{figure}
\includegraphics{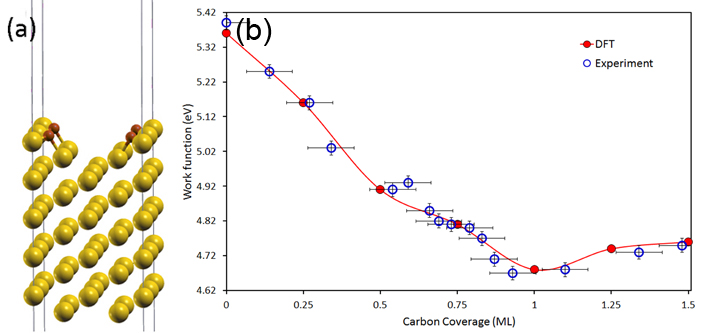}
\caption{\label{fig:workfunction}(a) Schematic of a ($2\times 2$) simulation super-cell which illustrates the missing row surface structure of Au(110)-(2$\times$1). This simulation box consists of 38 gold atoms (10 layers) with four adsorbed carbon atoms (corresponding to the 1 mono-layer coverage). The periodic slabs are separated by a  vacuum region. (b) Work function versus coverage for carbon atom adsorption on Au(110)-(2$\times$1). The DFT computational values are connected by cubic-spline interpolation. The error bars are shown for the experimental values.}
\end{figure}

Fig.~\ref{fig:workfunction}(a) shows the schematic of a ($2\times 2$) simulation super-cell which illustrates the missing row surface structure of Au(110)-$(2\times 1$). This simulation box consists of 38 gold atoms (10 layers) with four adsorbed carbon atoms (corresponding to one ML coverage). The periodic slabs are separated by a large 15\text{\AA} vacuum region. In this configuration, the carbon atoms are adsorbed on the pseudothreefold sites. This corresponds to the work function calculations at one ML coverage, ($\theta=1$), presented in Fig.~\ref{fig:workfunction}(b). 

The first prediction for the reduction of work function with deposition coverage came from Topping \cite{Topping1927}.  This model assumes that the dipoles are arranged as a planar network, and the distance between the dipoles decreases gradually with increasing the deposition coverage. Since the Topping model assumes the same mode of packing for all coverages, it only predicts a smooth dependence of the work function on the adsorbate coverage. Later, Gyftopoulos and Levine \cite{Gyftopoulos1962} developed a model which treats the adsorbate-induced work function as a simple sum of a dipole barrier and an electronegativity barrier. This model predicts that the degree of electron transfer from the adsorbed atom to the metal surface is proportional to the difference in electronegativities of the adsorbate and the substrate. Carbon atoms are only slightly more electronegative than the gold substrate (2.55 for C, and 2.54 for Au). Therefore the deposited carbon adatoms on the gold surface would be slightly polarized negatively outward, which would lead to an increase of the work function. 
This is, however, in opposition to the behavior exhibited in Fig.~\ref{fig:workfunction}(b). Starting from around 5.39 eV at zero coverage, the work function near-linearly decreases by the coverage. The plot, then, flattens off toward a minimum value of 4.68 eV at one mono-layer coverage. After that, the work function rises approximately to a value at 1.5 ML associated with the work function of the bulk adsorbate, \textit{i.e.} graphite.

\begin{figure}
\includegraphics[width=1.0\linewidth, height=12cm]{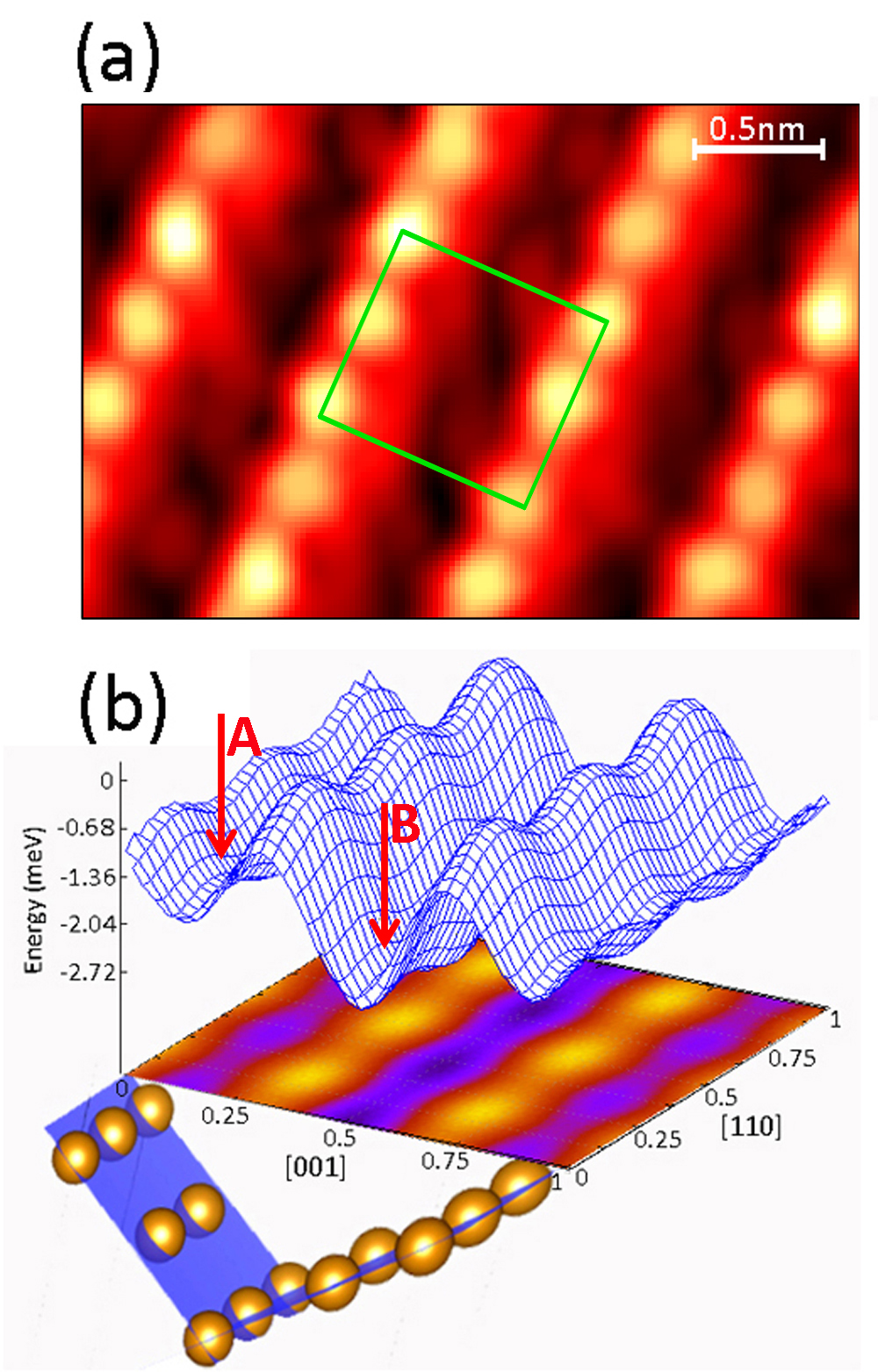}
\caption{\label{fig:STM} (a) STM image $3\times 2$ $nm^2$ of Au(110)-(2$\times$1). The four bright lines show the gold atoms, closed packed along the [110] direction. The three dark channels are the missing rows. The green box shows the surface edges of the periodic boundary condition used in simulations in this study. (b) At bottom, the atomic configuration of the top three layers of the missing row structure in Au(110)-(2$\times$1) surface is given. The potential energy landscape and its contour plot are presented on top. The outer and interior potential wells formed on this surface are indicated by red arrows and labeled by 'A' and 'B', respectively.}
\end{figure}

\begin{figure}
\includegraphics[width=0.7\linewidth, height=5cm]{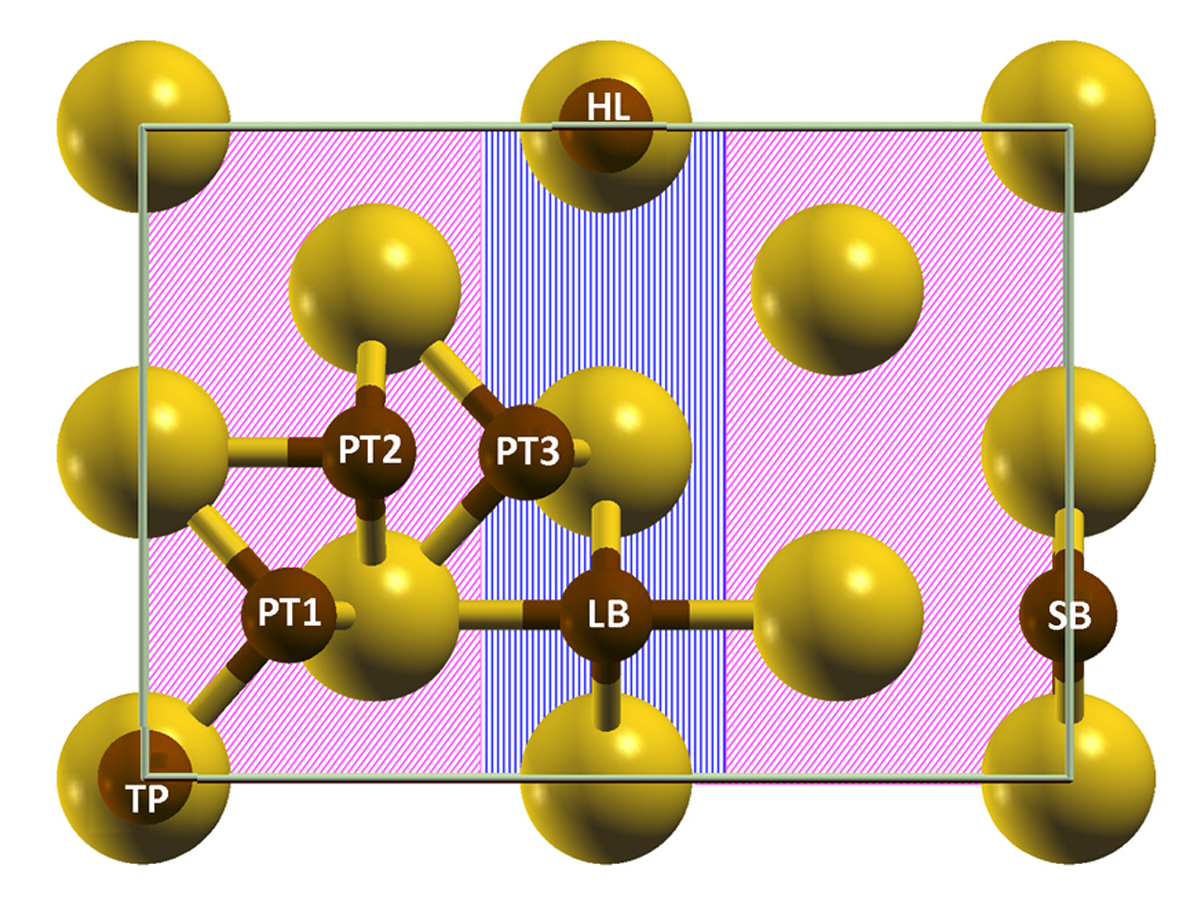}
\caption{\label{fig:Adsorption_Sites} (a) (Top view) There are seven adsorption sites on the Au(110)-(2$\times$1) surface \cite{Landmann2009}. The top (TP) and the short bridge (SB) sites are the outermost sites, which only serve as temporary hosts for the carbon adatoms during the deposition. The two pseudothreefold sites, PT1, and PT2 are immediately accessible local minima within the outer potential well of the trench (see also Fig.~\ref{fig:STM}(b)). With a lower probability, the carbon atoms may form another pseudothreefold configuration within the interior potential well, PT3. Adsorption of carbon atoms at the two central sites of the Au(110)-(2$\times$1), namely, the hollow (HL) and the long-bridge (LB) are also less likely. The red and blue dashed areas indicate the initial positioning of the carbon adatoms on top of the trench for the \textit{ab-initio} MD simulations (see the text).}
\end{figure}

In the remainder of this article, we explain the measured and calculated behavior of the work function with the carbon coverage by providing a dynamical mechanism for the subsequent adsorption of the carbon adatoms on the gold surface.
Fig.~\ref{fig:STM}(a) provides a high resolution STM image of the ($2\times 1$) reconstructed gold surface. The closed packed gold atoms along the [1$\bar{1}$0] direction are easily recognized in this image as four bright lines. The three dark channels are the missing rows. The green box shows the surface edges of the periodic boundary condition used in the DFT simulations. When a carbon atom approaches this surface, it is exposed to the potential energy surface (PES) shown in Fig.~\ref{fig:STM}(b). This potential surface is obtained by post processing the SCF results of the relaxed surface using the QE package. For clarity, the atomic configuration of the top three layers of the missing row structure in the surface is illustrated at the bottom of this figure. The contour plot corresponding to this potential energy landscape is also given in Fig.~\ref{fig:STM}(b). As indicated by red arrows, there are two (by symmetry) potential wells formed on this surface. As will be explained later, the outer potential well, labeled by 'A', provides local minimum sites for the deposited carbons. These sites would be immediately accessible for the adatoms compared to the more energetically favorable sites in the interior potential channel, labeled by 'B'.   
There are seven adsorption sites available on Au(110)-(2$\times$1) for the adatoms to occupy \cite{Landmann2009}. These sites are shown in Fig.~\ref{fig:Adsorption_Sites}. The top site,TP, is the least favorable, with E=4.52 eV with C-Au bond of 1.83 \text{\AA}. The twofold small bridge, SB, has an  adsorption energy of E=5.44 eV, with C-Au bond length of 1.90 \text{\AA}. The carbon atom at the hollow site, HL, makes a fivefold coordination with E=5.98 eV, and a C-Au bond length of 2.19-2.32 \text{\AA}. The most stable adsorption site is the long-bridge site, E=6.62 eV, with fourfold coordinated C, and bond lengths of 2.03 \text{\AA} and 2.10 \text{\AA}. As can be seen in the PES in Fig.~\ref{fig:STM}(b), the first two pseudothreefold sites, PT1 and PT2, are located in the outer local minimum channel formed between the first two top Au layers. The adsorption energies corresponding to these sites are E=6.28 eV, and E=6.14 eV, respectively. The third pseudothreefold site,PT3, E=5.97 eV, is formed between the second and third Au layers inside the main potential channel. 

\begin{figure}
\includegraphics[width=1.0\linewidth, height=9cm]{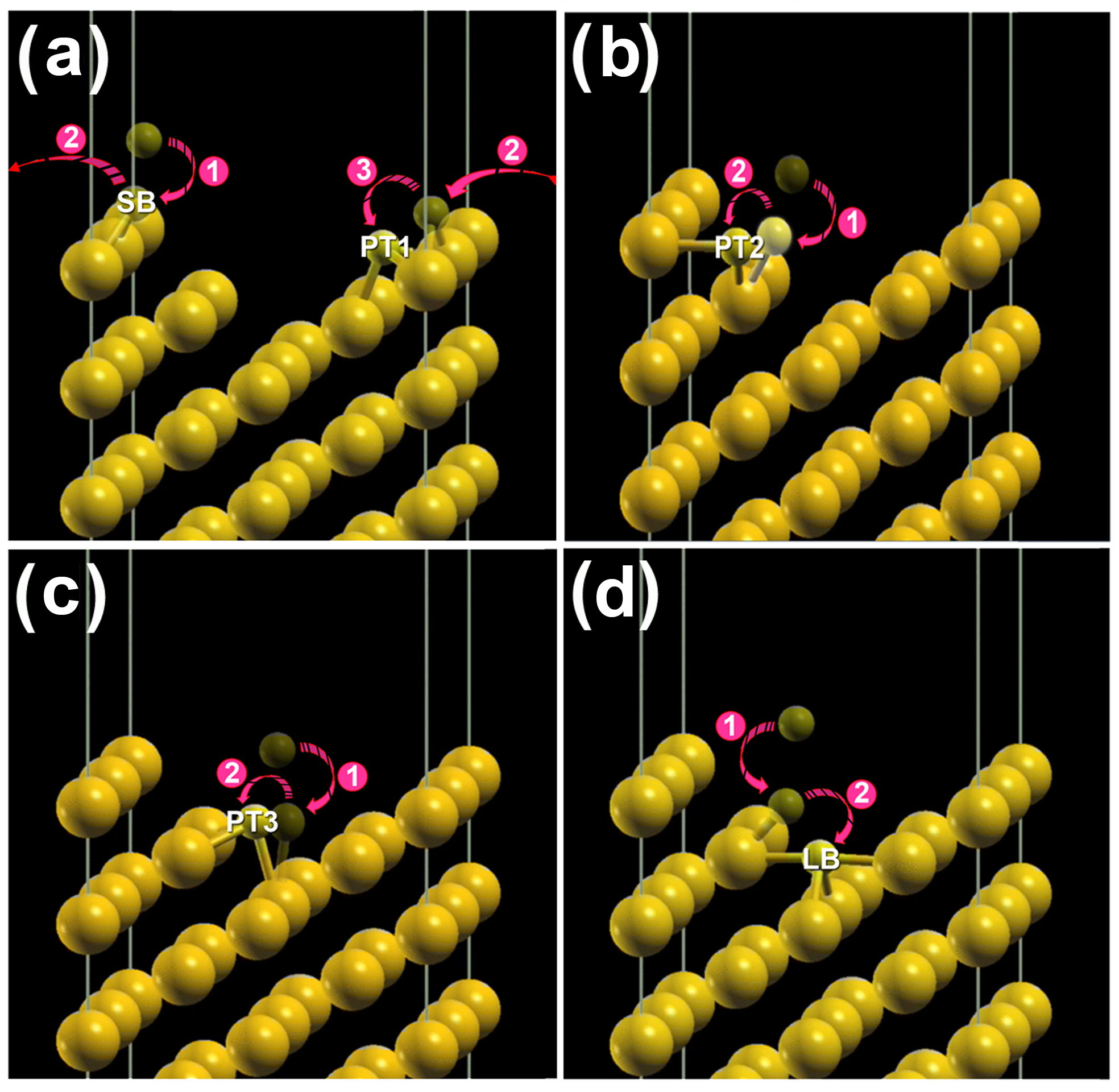}
\caption{\label{fig:MD_Snap_Shots} Snap shots of the MD simulations \cite{Movies}. In each case, the steps that the  adatom takes to relax to its final position are labeled numerically and indicated by red arrows. (a) Snap shot from Mov.(1). The carbon atom is initially placed on or around one of the top positions. It quickly moves toward the adjacent gold atom and forms the SB configuration, step (1). The jump across the periodic boundary is indicated by step (2). The carbon dynamics continues by transferring to the PT1 site, step (3). (b) Snap shot from Mov.(2), which shows the initial adsorption of a carbon adatom to an atop position, and its transition to a PT2 site. (c) Snap shot from Mov.(3). This figure shows that the carbon atom may land on a HL site, but it eventually relaxes to the more stable PT3 configuration. (d) Snap shot from Mov.(4), which shows the transition of a carbon adatom from being adsorbed on a single gold atom to the most energetically favorable site, $\textit{i.e.}$ LB.}
\end{figure}

As we will show next, by {\it ab-initio} molecular dynamics (AIMD) calculations, the two immediately accessible sites, TP and SB, serve as intermediate hosts for the adsorbate adatoms; meaning that although the carbon atoms may initially attach to these sites, within one picosecond time interval, they move and relax down toward the more stable PT2, and PT1 positions, respectively. Whether by direct adsorption on PT1 and PT2 sites, or transferring from TP and SB sites, when the carbon adatoms are trapped in these local minima, it is less likely for them to overcome the existing potential barrier and reach the most stable site, LB, see the PES in Fig.~\ref{fig:STM}(b).  
The first movie, Mov.(1), \cite{Movies}, in the supplemental material (SM), shows 1 ps of the MD simulation when a carbon atom is initially placed on or around one of the top positions. As can be seen in this movie, the carbon atom quickly moves toward the adjacent gold atom and forms the SB within the first 0.1 ps of the simulation. The carbon dynamics continues by transferring to the PT1 site after about 0.4 ps. The carbon atom holds this position for the rest of the simulation time. The snap shot of the Mov.(1) is presented in Fig.~\ref{fig:MD_Snap_Shots}(a). As will be discussed later, the subsequent addition of more adatoms to the system does not push the first carbon out of its PT1 site, primarily because of the large barrier between the 'A' and 'B' potential wells. As our further MD simulations demonstrate, positioning the carbon atom on top of the supercell within the red-dashed area in Fig.~\ref{fig:Adsorption_Sites} results in the adsorption of the carbon atom in PT1 or PT2 sites (see Mov.(1) and Mov.(2) in the supplemental material \cite{Movies}). Initial adsorption of carbon adatom to a atop site and its transition to a PT2 site is labeled and indicated by red arrows in Fig.~\ref{fig:MD_Snap_Shots}(b). 

As another scenario, if a carbon atom approaches the surface within the region indicated by the narrow blue dashed area in Fig.~\ref{fig:Adsorption_Sites}, it will eventually be trapped in one of the HL or LB sites at the interior part of the Au(110)-(2$\times$1) trench (see Mov.(3) and Mov.(4) \cite{Movies}, and the corresponding snap shots in Figs.~\ref{fig:MD_Snap_Shots}(c) and (d), respectively). As can be seen in Mov.(3), the HL site may also serve as an intermediate stage for the carbon atom before it eventually lands on the most stable site, \textit{i.e.} LB. The individual MD simulations with various initial start positions for carbon atoms and the size comparison between the red and blue areas in Fig.~\ref{fig:Adsorption_Sites} reveal that a carbon atom would have a higher chance to get trapped in the outermost adsorption sites, rather than to penetrate deeper into the Au(110)-(2$\times$1) trench and reach the more energetically favorable position.
The domination of the side adsorption, \textit{i.e.} within the 'A' potential well, is also manifested in the measured work fucntion; as the work function decreases by deposition of the first carbon atom (0.25 ML experimental coverage corresponds to the adsorption of one adatom on the given (2$\times$2) simulation supercell). The correlation between the work function reduction and the adsorption sites is provided by analyzing the electron charge density distribution before and after carbon adsorption at different sites on Au(110)-(2$\times$1).

\begin{figure}
\includegraphics{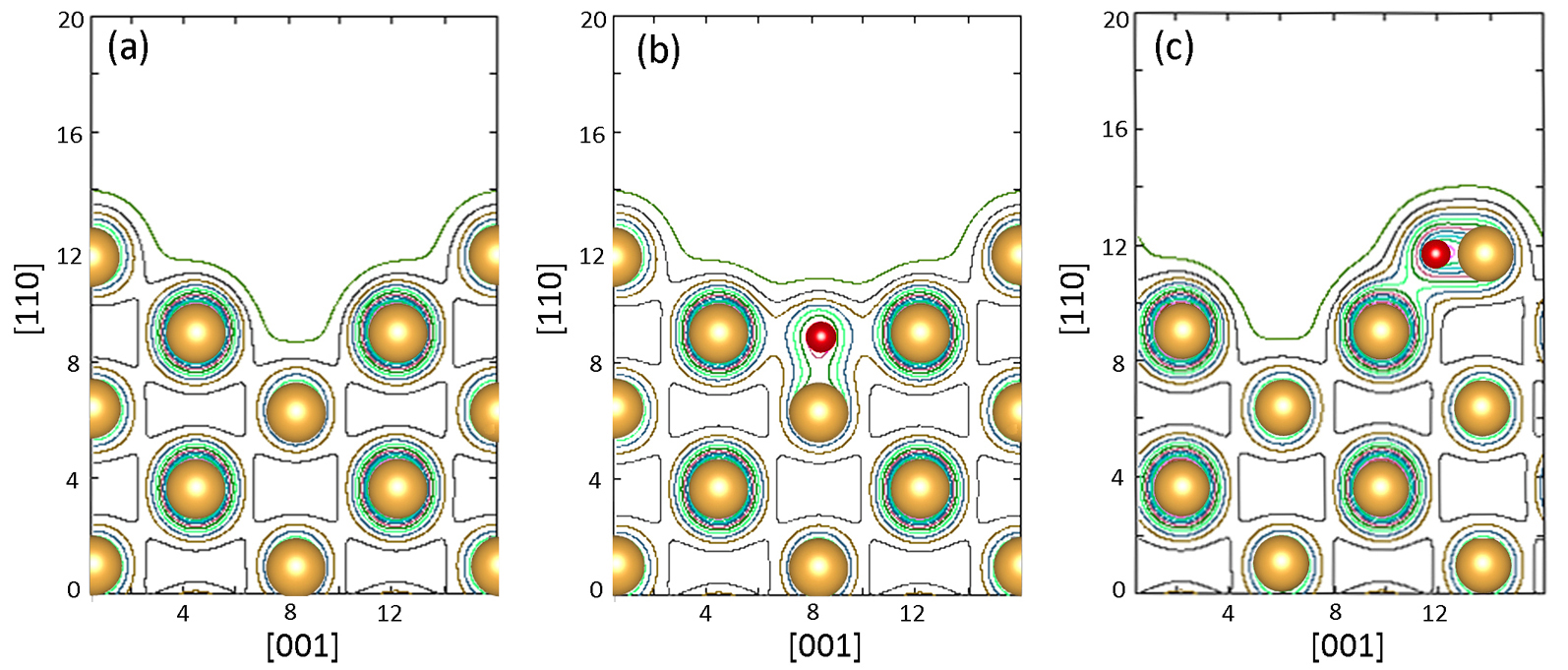}
\caption{\label{fig:Charge} Electron charge density distribution for (a) clean  surface. The roughness of the missing row structure makes the effective role of smoothing due to the adsorption more pronounced. (b) A carbon atom is adsorbed on the long bridge (LB) site. The adsorption of carbon atom (represented in red color) into the rough Au(110)-(2$\times$1) surface leads to the smoothing of the electronic charge distribution of the surface and hence to the enhancement of the work function. (c) A carbon atom is adsorbed on the PT1 site between the first two top layers of gold. The positioning of the carbon atom on the upper side of trench gives rise to a dipole with the positive side outwards and hence lowers the work function.}
\end{figure}

The electron charge density plots presented in Fig.~\ref{fig:Charge} are obtained by post processing the SCF results of the relaxed surface. As can be seen in Fig.~\ref{fig:Charge}(a), the roughness of the missing row structure of the Au(110)-(2$\times$1) surface makes the effective role of the carbon adsorption on the electron charge density distribution more pronounced. As illustrated in Fig.~\ref{fig:Charge}(b), if the carbon atom can penetrate deep into the trench, its adsorption on the LB site would lead to a significant smoothing of the electron charge density distribution. 

\begin{figure}
\includegraphics{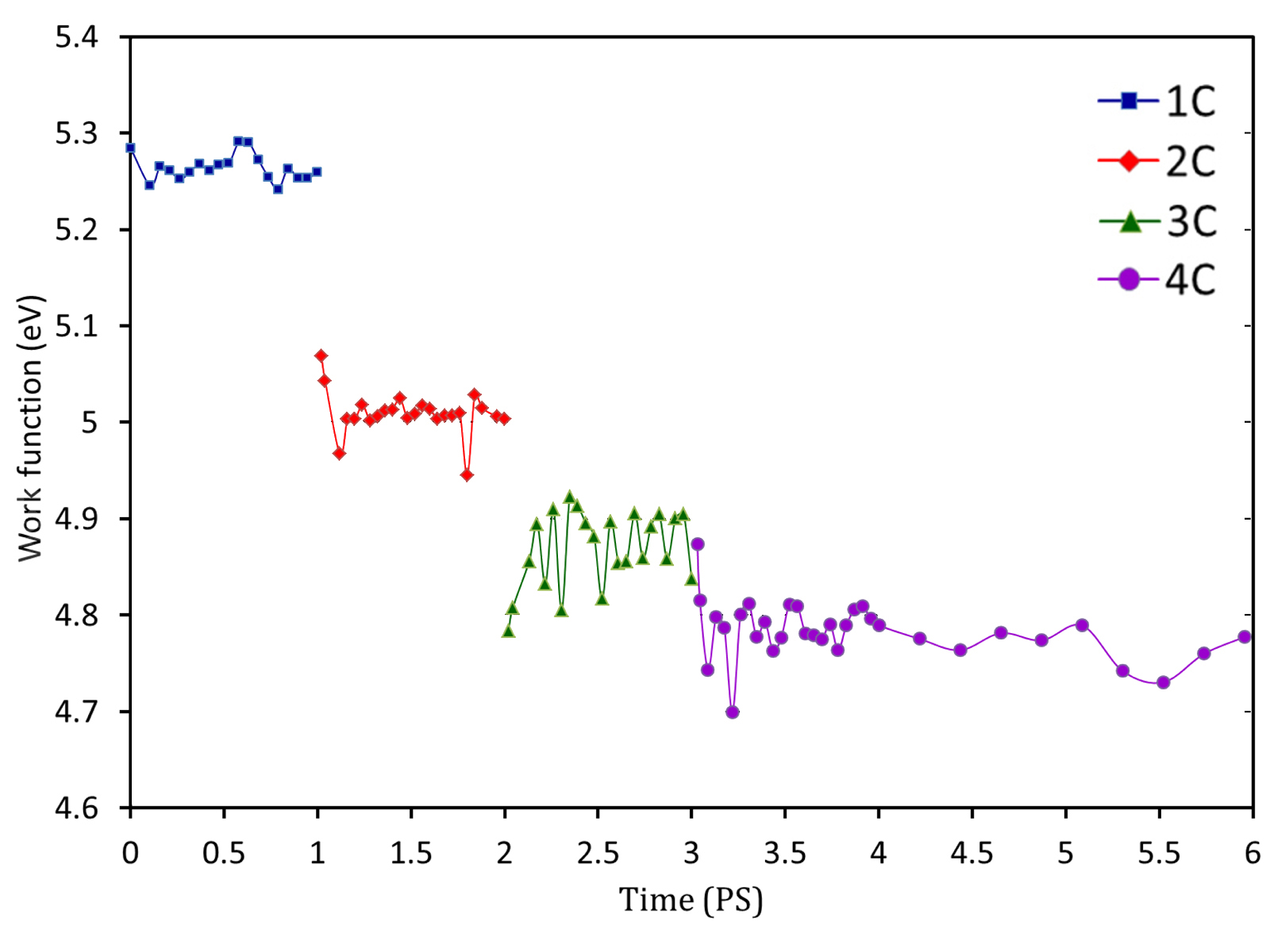}
\caption{\label{fig:MD_Deposition} Work function versus time within 6 ps of \textit{ab-initio} molecular dynamics simulations. The work function calculations are done at 50 fs time steps. Compared to Fig.~\ref{fig:workfunction}(b), the slightly enhanced values for the work function at each coverage can be explained by the lower level of theory implemented in the MD simulation.}
\end{figure}

This smoothing effect has previously been studied \cite{Rhead1988}, increasing the work function with a calculated value of \textit{W}=5.44 eV. This is contrary to the observed decreasing pattern for the work fucntion at the initial coverage values. However, as shown in Fig.~\ref{fig:Charge}(c), the adsorption of the carbon atom on the PT1 site makes the Au(110)-(2$\times$1) rough surface even rougher. It is also known \cite{Muscat1986,Serena1987} that this effect lowers the work function giving rise to a dipole with the positive side outwards.
We now consider the work function changes over time when carbon atoms are added to the Au(110)-(2$\times$1) surface subsequently until a ML coverage is achieved (\textit{i.e.} four carbon atoms within the (2$\times$2) simulation supercell). This result is presented in Fig.~\ref{fig:MD_Deposition}. Compared to Fig.~\ref{fig:workfunction}(b), the slightly enhanced values for the work function at each coverage can be explained by the lower level of theory implemented in the MD simulation. As initially shown in Fig.~\ref{fig:workfunction}(a), the final configuration will have the four carbon atoms occupying the four available PT1 sites in the simulation cell. 
Mov.(5) \cite{Movies} in the SM shows 6 ps of MD simulations for this system. When a carbon atom is added to the system, it eventually gets to a PT1 position (passing through a SB site). Also, the addition of further carbon atoms do not push the adatoms away from their PT1 sites. The fluctuation of the work function due to the oscillations of the carbon atoms around its equilibrium position at the adsorbed sites are observable in Fig.~\ref{fig:MD_Deposition}. The work function calculations are done at 50 fs time steps at each of the 1 ps MD simulations for subsequent carbon additions. The work function calculations for the final configuration is reported for the extended time (2 ps) to achieve the convergence.
\section{Summary and outlook}
In this work, we studied the mechanism of work function variation with carbon coverage on reconstructed Au surfaces. The calculated work function with carbon adatom coverage was compared with measurements. The surface dynamics of carbon adsorption was investigated using {\it ab initio} molecular dynamics; and the contribution of various available adsorption sites on the observed work function content of the surface is determined. The surface hopping from the intermediate adsorption sites to the more energetically favorable positions were studied up to a mono-layer coverage of carbon adatoms on the Au(110)-(2$\times$1) missing row structure. In the future, using long-time AIMD simulations with high concentrations of carbon adatoms, we aim to obtain the diffusion constant and the transition rates between different diffusion paths, with temperature. Such time domain calculations, will provide us with the determination of the diffusion constant which is not yet accurately available in Eq.~\ref{eqn_1}.

\section{\label{sec:level1} acknowledgments}
 HZJ is supported by a theory grant from the NIST Physical Measurement Laboratory. HRS acknowledges support from the NSF through a grant for ITAMP at Harvard  University. Sandia National Laboratories is a multi-mission laboratory managed and operated by National Technology and Engineering Solutions of Sandia, LLC., a wholly owned subsidiary of Honeywell International, Inc., for the U.S. Department of Energy’s National Nuclear Security Administration under contract DE-NA0003525. The views expressed in the article do not necessarily represent the views of the U.S. Department of Energy. This article is a contribution of the U.S. Government and is not subject to U.S. copyright. 

\newpage
\bibliography{references}

\end{document}